\documentclass[12pt,twoside]{article}
\usepackage{fleqn,espcrc1,epsfig}
\input BoxedEPS
\SetTexturesEPSFSpecial

\usepackage{graphicx}

\newcommand{\AmS}{{\protect\the\textfont2
   A\kern-.1667em\lower.5ex\hbox{M}\kern-.125emS}}

\title{
A Relativistic Symmetry in Nuclei: Its origins and consequences}
\author{
Joseph N. Ginocchio\\
MS B283, Los Alamos National Laboratory, Los Alamos, NM, 87545,USA }
\begin{document}
\maketitle
\begin{abstract}
We review the status of quasi-degenerate doublets in nuclei, called 
pseudospin doublets, which were discovered about thirty years ago
and the origins of which have remained a mystery, until recently.  We 
show that pseudospin doublets originate from an SU(2) symmetry of
the Dirac Hamiltonian which occurs when the sum of the scalar and vector 
potentials is a constant. Furthermore, 
we survey the evidence that pseudospin symmetry is approximately 
conserved in nuclear spectra and eigenfunctions and in nucleon-nucleus scattering for a Dirac Hamiltonian with realistic nuclear
scalar and vector potentials.
\end{abstract}
\section{INTRODUCTION}
\label{sec:intro}
I first met Achim Richter at an international nuclear physics 
conference in Crete in 1982. We talked about the 
"scissors" mode, an excited state with
quantum numbers
$J^{\pi} = 1^+$, which he and his collabrators had discovered 
experimentally at 3.075 MeV in $^{156}Gd$
\cite {achim}. My collaborators and I had predicted in a schematic 
shell model calculation a $1^+$ in the
nearby Samarium isotopes at 3.0 - 3.4 MeV (depending on the isotope)
\cite{yoshida}. No one paid any attention to that calculation. I told 
Achim about our calculation at the
meeting, and, to his credit, he did acknowledge our work although 
everybody else continued to ignore our
prediction. Years later during a four month sabbatical to the Max Planck Institute in 
Heidelberg on a Humboldt Senior Scientist
Award, I visited the Technical University at Darmstadt for one 
day. I was very impressed with the facility
and the group that Achim was leading. He and his colleagues told me 
about the interesting work they were doing
which included by then a very exhaustive study of the ``scissors'' 
mode and its excitation strength in many
nuclei. The visit must have been very inspirational because the next 
day after returning to Heidelberg I
derived (within the interacting boson model
\cite {iachello}) a sum rule which related the total ``scissors'' 
excitation strength  to the number of quadrupole pairs of nucleons in 
the ground state \cite
{gino1}. Later on Achim, P. von Neumann-Cosel, H. Bauer, and I 
collaborated on a paper which used this sum rule to correlate the 
``scissors'' mode excitation
strength to the nuclear deformation \cite{cosel}. My work on the 
``scissors mode'' was very exciting for me and I am glad that Achim
persisted and found that little $1^+$ state.

In this paper I shall talk about quasi-degenerate doublets discovered 
more than thirty years ago \cite {aa,kth} called pseudospin symmetry 
doublets.  One of the results I shall show is
that this symmetry is related to another interest of Achim, namely 
the measurement of ``$\ell$'' forbidden magnetic dipole transtions 
\cite {achim2}.

\section{PSEUDOSPIN SYMMETRY}
The spherical shell model orbitals that were observed to be 
quasi-degenerate have non -
relativistic quantum numbers ($n_r$,
$\ell$, $j =
\ell + 1/2)$ and ($n_{r}-1, \ell + 2$, $j^{\prime} = j + 1 = \ell + 3/2$) 
where $n_r$, $\ell$, and $j$ are
the single-nucleon radial, orbital, and total
angular momentum quantum numbers, respectively
\cite {aa,kth}. This doublet structure is expressed in
terms of a ``pseudo'' orbital angular momentum
$\tilde{\ell}$ = $\ell$ + 1, the average of the orbital angular 
momentum of the two
states in doublet, and ``pseudo'' spin, $\tilde s$ = 1/2.
For example,
$(n_r s_{1/2},(n_r-1) d_{3/2})$ will have
$\tilde{\ell}= 1$ , $(n_r p_{3/2},(n_r-1) f_{5/2})$ will have 
$\tilde{\ell}= 2$, etc. Since $j = \tilde{\ell}\ - {1\over 2}$ and $j^{\prime} = \tilde{\ell}\ 
+ {1\over 2}$, the 
energy of the two states
in the doublet are then appproximately independent of the orientation 
of the pseudospin. Some examples are given in Table 1.\\
\begin{table}[h]
\noindent Table 1\\
Experimental (Exp) and relativistic mean field (RMF) \cite{mad2}
pseudospin binding energy
splittings $\epsilon_{j^{\prime}= \tilde{\ell}+1/2}-\epsilon_{j=
\tilde{\ell}-1/2}$ for various doublets in $^{208}Pb$.
\begin{center}
\begin{tabular}{cccc}

\hline

   $\tilde{\ell}$ & ($n_{r}-1, \ell + 2$, $j^{\prime} = \ell + 3/2$)
&
$\epsilon_{j^{\prime}= \tilde{\ell}+1/2}-\epsilon_{j= \tilde{\ell}-1/2}$ &
$\epsilon_{j^{\prime}=
\tilde{\ell}+1/2}-\epsilon_{j= \tilde{\ell}-1/2}$  \\
         &-($n_r$,
$\ell$, $j =
\ell + 1/2)$    & (Exp) & (RMF)  \\
        &    & (MeV) & (MeV)  \\
\hline

  4 &  $0h_{9/2}-1f_{7/2}$  &  1.073 &    2.575   \\
3 &  $0g_{7/2}-1d_{5/2}$  &  1.791   &  4.333   \\
   2 &  $1f_{5/2}-2p_{3/2}$  & -0.328  &   0.697 \\
1 &  $1d_{3/2}-2s_{1/2}$  &  0.351   &  1.247     \\
\hline
\end{tabular}\\
\end{center}
\end{table}\\

Pseudospin ``symmetry'' was shown to exist in deformed nuclei as well 
\cite {bohr} and has been used to explain
features of deformed nuclei, including superdeformation and identical bands
\cite{twin,stephens,stephens2,von}, and example of which is 
given in Fig. 1.  \\
\begin{figure}[h]
\begin{center}
\HideDisplacementBoxes
\BoxedEPSF{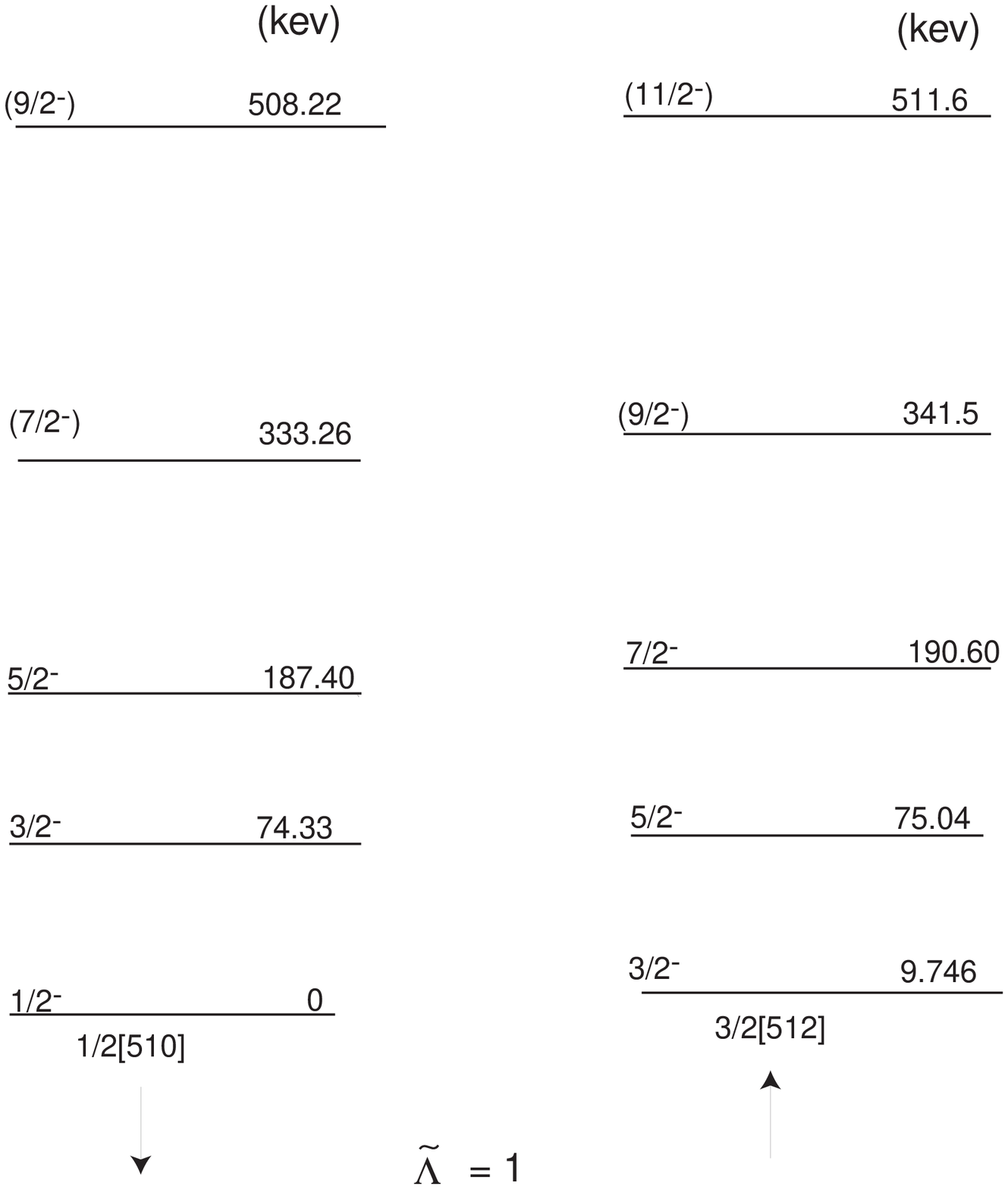 scaled 500}
\end{center}
\noindent Figure 1. Two quasi-degenerate rotational bands in 
$^{187}Os$, one with pseudospin unaligned with
the body-fixed pseudo-angular momentum projection ${\tilde \Lambda} 
=1$ and one aligned \cite {stephens2,von}.\\
\end{figure}

However, the origin of pseudospin symmetry remained a mystery and 
``no deeper understanding of
the  origin of
these (approximate) degeneracies'' existed
\cite {ben}. In this paper we shall review more recent
developements that show that pseudospin symmetry is a relativistic 
symmetry \cite
{gino,ami,gino2}.

\section{SYMMETRIES OF THE DIRAC HAMILTONIAN}

The success of the shell model implies that nucleons move in a mean field
produced by the interactions between the nucleons.
Normally, it suffices to use the Schrodinger equation to describe the 
motion of the
nucleons in this mean field.  However, in order to
understand the origin of pseudospin symmetry, we need to take into account the
motion of the nucleons in a relativistic mean field and
thus use the Dirac equation. The Dirac Hamiltonian, $H$, with an 
external scalar, $V_S$, and
vector, $V_V$, potentials is given by:
\begin{equation}
H =  {\vec{\alpha}} \cdot {\vec p} + \beta (M + V_S) + V_V ,
\label {dirac}
\end{equation}
where we have set
$\hbar = c =1$,  ${\vec \alpha}$, $\beta $ are the usual Dirac 
matrices, $M$ is the nucleon mass, and $\vec p$ is the three momentum. The Dirac Hamiltonian 
is invariant under an SU(2) algebra 
for two limits:
$V_S - V_V$ = constant and $V_S + V_V$ = constant \cite{bell}. The 
first condition leads to a spin symmetry which is relevant for
mesons \cite{bell,page} while the second leads to pseudospin symmetry 
\cite {ami}.
The generators for the SU(2) pseudospin symmetry,
${\hat {\tilde S}}_i$, which commute with the Dirac Hamiltonian, 
$[\,H\,,\, {\hat {\tilde S}}_i\,] = 0$,
are given by
${\hat {\tilde S}}_i = \left ( {{\hat {\tilde s}_i} \atop 0 } { 0 
\atop { {\hat { s}}_i}}\right )$ where ${{\hat { s}}_i}$ is the usual 
spin
operator,
${{\hat {\tilde s}}_i} = U_p\ {\hat {s}}_i \ U_p$ and $U_p = \, 2 {{ 
\hat { s}\cdot \vec p} \over p}$ is a unitary
helicity operator \cite {jerry}. These pseudospin generators have the spin operator 
${\hat {s}}_i$ operating on the lower component
of the Dirac wave function which has the consequence that the spatial 
wavefunctions for the two states in the
pseudospin doublet are identical to within an overall phase.

This symmetry for $V_S + V_V$ = constant is general and applies to 
spherical, axially deformed, triaxial, gamma unstable, etc., 
nuclei. In the case for which the potentials are spherically 
symmetric, the Dirac Hamiltonian
conserves the pseudo-orbital angular momentum, the generators of which are, $
{\hat {\tilde L}}_i =
\left ( {{\hat {\tilde \ell}_i} \atop 0 } { 0 \atop { {\hat { 
\ell}}_i}}\right ),$
where ${\hat {\tilde \ell}}_i = U_p\, {\hat \ell}_i$ $U_p$, 
${\hat\ell}_i = \vec r \times
\vec p$. Since $U_p$
conserves the total angular momentum but $\vec p$ changes the orbital angular
momentum by one unit because of parity conservation, if the lower 
component of the Dirac wave function orbital angular momentum
is ${\tilde \ell}$, the upper component also has total angular momentum $j$,
but orbital angular momentum $\ell = {\tilde \ell} \pm
1$. If $j = {\tilde \ell} + 1/2$, then it follows that $\ell = 
{\tilde \ell} + 1$, whereas if $j = {\tilde \ell} - 1/2$, then $\ell 
= {\tilde \ell} -
1$. This agrees with the pseudospin doublets originally observed 
\cite{aa,kth} and discussed at
the beginning of this paper. This relativistic interpretation also 
gives the physical significance of the pseudo-orbital angular 
momentum ${\tilde
\ell}$ as the ``orbital angular momentum'' of the lower component. 

For axially symmetric deformed nuclei, there is
a U(1) generator corresponding to the pseudo-orbital angular momentum
projection along the body-fixed symmetry axis which is conserved in addition to the
pseudospin, ${\hat{\tilde \lambda}} = \left ( {{\hat {\tilde 
\Lambda}} \atop 0 } { 0 \atop {{\hat \Lambda}} }
\right ),$ where ${\hat {\tilde \Lambda}} = U_p\ \hat \Lambda\ U_p$.

In general, the eigenfunctions of the Dirac Hamiltonian,
$H \Psi_{\tau,\tilde\mu} = {\cal E}_{\tau} \Psi_{\tau,\tilde\mu}$, are
doublets (${\tilde S} =1/2$,
$\tilde\mu =\pm 1/2$) with respect to the SU(2) generators ${\hat 
{\tilde S}}_i$ 
\begin{equation}
{\hat {\tilde S}}_z\,\Psi_{\tau,\tilde \mu} = \tilde\mu\, 
\Psi_{\tau,\tilde\mu} ~, \nonumber\\
{\hat {\tilde S}}_{\pm}\,\Psi_{\tau,\tilde \mu} = {\sqrt{(1/2 \mp 
{{\tilde \mu}})( 3/2 \pm {{\tilde \mu}}) }}\, \Psi_{{\tau,
\tilde \mu}
\pm 1} ~,
\label {ugen}
\end{equation}
where ${\hat {\tilde S}}_{\pm} = {\hat {\tilde S}}_{x} \pm i{\hat 
{\tilde S}}_y$. The eigenvalue $\tau$ refers to the other necessary
quantum numbers.

However, the exact symmetry limit can not be realized in nuclei, 
because, if $V_S + V_V$ = constant,
there are no Dirac bound valence states and hence nuclei
can not exist \cite{su,gino}.

\section{REALISTIC RELATIVISTIC MEAN FIELDS}

A near equality in the magnitude
of mean fields, $V_S \approx - V_V$, is a universal feature of the 
relativistic mean field
approximation (RMF) of
relativistic field theories with interacting nucleons and mesons 
\cite {wal} and relativistic
theories with nucleons interacting via zero range
interactions
\cite{mad}, as well as a consequence of QCD sum rules \cite{furn}.
Recently realistic
relativistic mean fields were shown to exhibit approximate pseudospin 
symmetry in both
the energy spectra and wave functions \cite {mad2,ring,arima}. In 
Table 1 pseudospin-orbit splittings calculated in the RMF \cite {mad2} are
compared with the measured values and are seen to be larger than the 
measured splittings which demonstrates that the pseudospin
symmetry is better conserved experimentally than mean field theory 
would suggest. As mentioned in the last section
pseudospin symmetry implies that the spatial wavefunction for the 
lower component of the Dirac wavefunctions will be
equal in shape and magnitude for the two states in the doublet. In 
Figure 2, for example, the lower components of the
$(2s_{1/2},1d_{3/2})$ Dirac eigenfunctions are approximately 
identical whereas the upper components have a different number of 
nodes
\cite {mad2}.  We also note that the lower components of the Dirac 
wavefunction are much smaller than the upper components which is 
consistent with the
view that nuclei are primarily non-relativistic. Nevertheless the 
understanding of the pseudospin symmetry involves
relativity.
\begin{figure}[h]
\begin{center}
\HideDisplacementBoxes
\BoxedEPSF{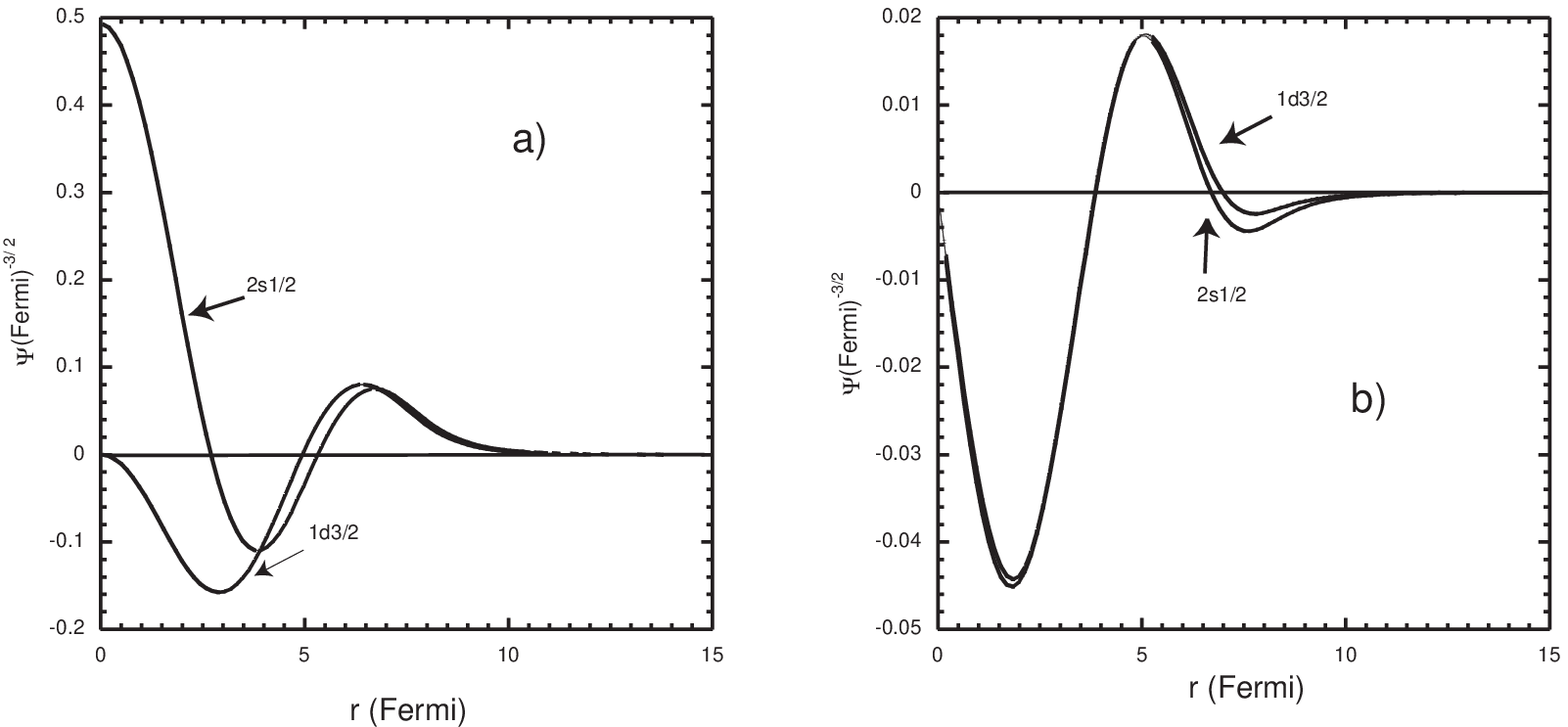 scaled 1000}
\end{center}
\noindent Figure 2. The upper a) and lower components b) of 
$(2s_{1/2},1d_{3/2})$ Dirac eigenfunctions in $^{208}Pb$ \cite 
{mad2}.\\
\end{figure}

\section{MAGNETIC DIPOLE AND GAMOW TELLER TRANSITIONS}
Because the lower components are small, in order to test the pseudospin 
symmetry prediction that the lower components are almost identical 
we must observe
transitions for which the upper components are not dominate. Magnetic 
dipole and Gamow-Teller transtions between the states in the
doublet are forbidden non-relativistically since the obital angular 
momentum of the two states differ by two units, but are allowed 
relativistically.
Pseudospin symmetry predicts that, if the magnetic moments, $\mu$, of the two 
states are known, the magnetic dipole transition,$B(M1)$, between the states 
can be
predicted. Likewise if the Gamow - Teller transitions between the 
states with the same quantum numbers are known, the transition 
between the states
with different quantum numbers can be predicted \cite {gino3}. For 
example for neutrons, the M1 transition is given
by
\begin{equation}
  \sqrt{B(M1:{n_{r}-1},{\ell}+2, j^{\prime} \rightarrow {\tilde
{n_{r}}}, {\ell}, j} )_{\nu} = - \sqrt{j + 1\over 2j + 1}( \mu_{j,\nu} -
\mu_{A,\nu}),
\label {neutron}
\end{equation}
\begin{equation}
  \sqrt{B(M1:{n_{r}-1}, {\ell}+2, j^{\prime} \rightarrow {\tilde
{n_r}}, {\ell}, j} )_{\nu} =  {j + 2 \over 2j + 3}\  \sqrt{2j +
1\over j + 1}(\mu_{j^{\prime},\nu} + {{j+1}\over {j + 2}} \mu_{A,\nu}),
\label {neutron2}
\end{equation}
where $j^{\prime} = {{\ell}} + 3/2, j = {{\ell}} + 1/2$ and $ 
\mu_{A,\nu} = -1.913 \mu_{0}$ is the anomolous magnetic moment. A 
survey of forbidden
magnetic dipole transitions taking into account the single-particle 
corrections by using spectroscopic factors shows a reasonable agreement 
with these
relations, an example of which is given in Figure 3 \cite{peter}.
\begin{figure}[h]
\begin{center}
\HideDisplacementBoxes
\BoxedEPSF{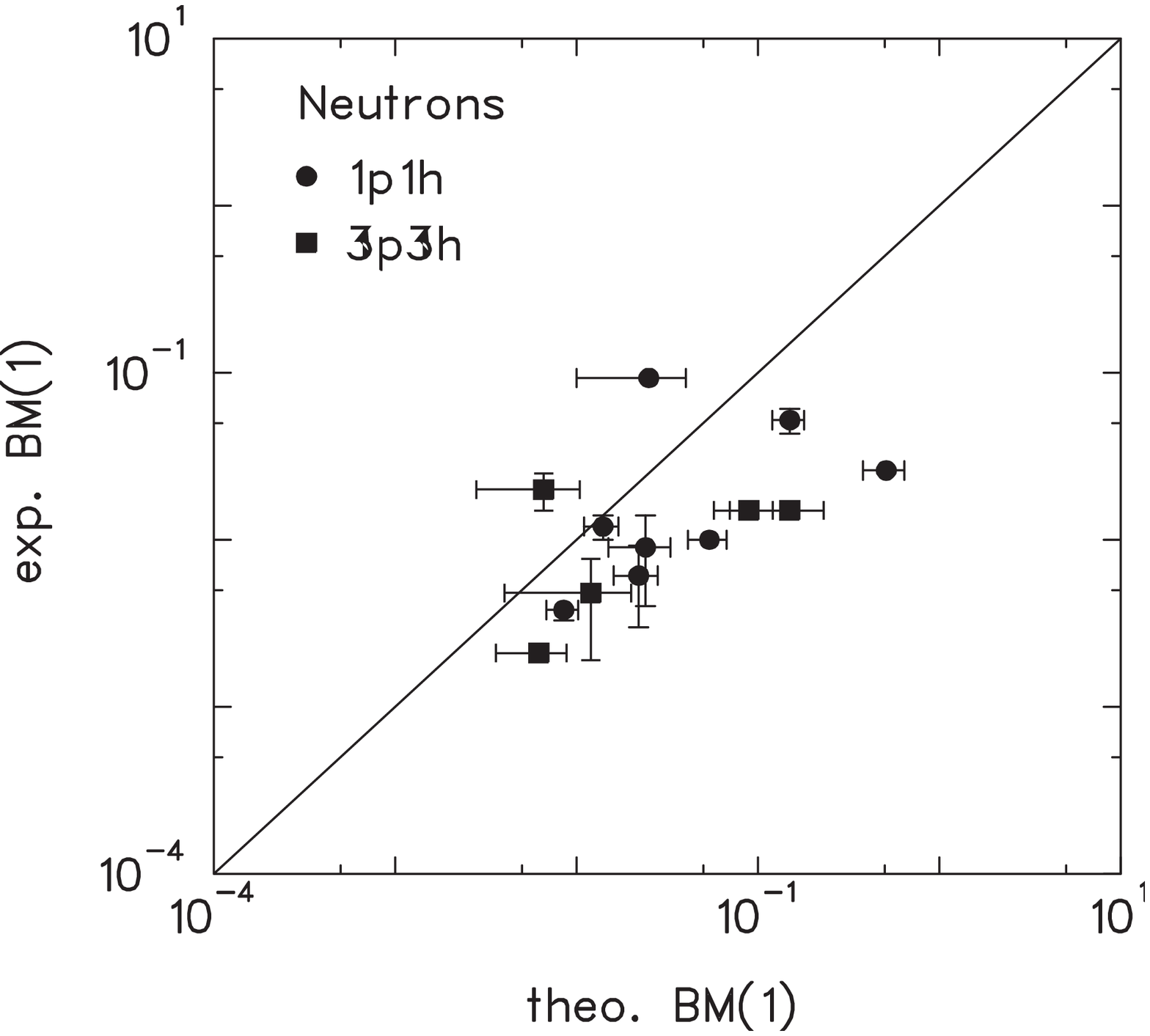 scaled 400}
\end{center}
\noindent Figure 3. The experimental and theoretical B(M1) between 
one-particle or one-hole (circles) states and
three-particle or three hole (squares) states in the doublet. The 
straight line denotes perfect agreement.\\
\end{figure}

\section{PSEUDOSPIN PARTNERS}

In the pseudospin symmetry limit, the eigenstates of the Dirac 
Hamiltonian in the doublet are connected by the pseudospin generators 
as given by (\ref
{ugen}).  For nuclei, pseudospin symmetry is broken and the 
pseudospin partner produced by the raising and lowering
operators on an eigenstate will not necessarily be an eigenstate. The 
question is: how different is the pseudospin
partner from the eigenstate with the same quantum numbers? Energy 
splittings suggest that the symmetry breaking is small (Table 1) but is
the breaking in the eigenfunctions small as well?  

The pseudospin 
generators operating on the spherical basis which has
pseudo-orbial angular momentum and spin coupled to total angular 
momentum $j$ and total angular momentum projection $m$ will give
\begin{equation}
S_q |n_r-1,\  \ell + 2,\  j^{\prime} ,\ m> = 
A_{j^{\prime},j^{\prime}}\ \ |n_r-1,\  \ell + 2,\
j^{\prime},\ m^{\prime}> + A_{j^{\prime}, j}\ |n_r,\  \ell,\  j,\ 
m^{\prime}>_{psp},
\end{equation}
\begin{equation}
  S_q |n_r,\  \ell,\  j ,\ m> = A_{j,j}\ |n_r,\  \ell,\  j,\ 
m^{\prime}> + A_{j, j^{\prime}}\ |n_r-1,\  \ell + 2,\  j^{\prime},\ 
m^{\prime}>_{psp},
\label {psp}
\end{equation}
where $A_{j, j^{\prime}} = (-1)^{{1\over2}-m^{\prime}+\tilde 
{\ell}} \sqrt{3(2j+1)(2j^{\prime}+1)\over 2} \left ( {j^{\prime} \atop -m^{\prime} } {
1
\atop q} { j \atop m}\right )
\
\left
\{ {{1\over 2} \atop j^{\prime} } { \tilde {\ell} \atop 1} { j \atop 
{1\over 2}}\right \}$. In the pseusospin limit
\begin{equation}
|n_r,\  \ell,\  j,\ m^{\prime}>_{psp}=|n_r,\
\ell,\  j,\ m^{\prime}>, \ V_S + V_V = constant,
\end{equation} 
\begin{equation}
|n_r-1,\  \ell + 2,\  j^{\prime},\ 
m^{\prime}>_{psp}=|n_r-1,\  \ell + 2,\ j^{\prime},\ m^{\prime}>, \ V_S + V_V = constant.
\end{equation} 
Since the lower components
of the eigenstates are operated on by the usual spin operator, ${\hat 
 s}_i$, the radial function of the
pseusospin partner is the same as the original eigenstate and hence 
Figure 2b shows how close the partner and
eigenstate are. For the upper component, the operator ${{\hat {\tilde 
s}}_i}$ intertwines space and spin degrees
of freedom through
$U_p = \, 2 {{  \hat { s}\cdot \vec p} \over p}$, thus ${{\hat {\tilde 
s}}_i}$ changes both the radial quantum number and the orbital angular 
momentum. In
Figure 4 we compare the spatial wavefunctions of the upper components of the pseudospin 
partner with the eigenstate with the same quantum numbers and we see that the wavefunctions are 
almost equal in the interior but differ significantly as the surface is apporached. These 
differences shall be discussed in more detail \cite {gino5}.
\begin{figure}[h]
\begin{center}
\HideDisplacementBoxes
\BoxedEPSF{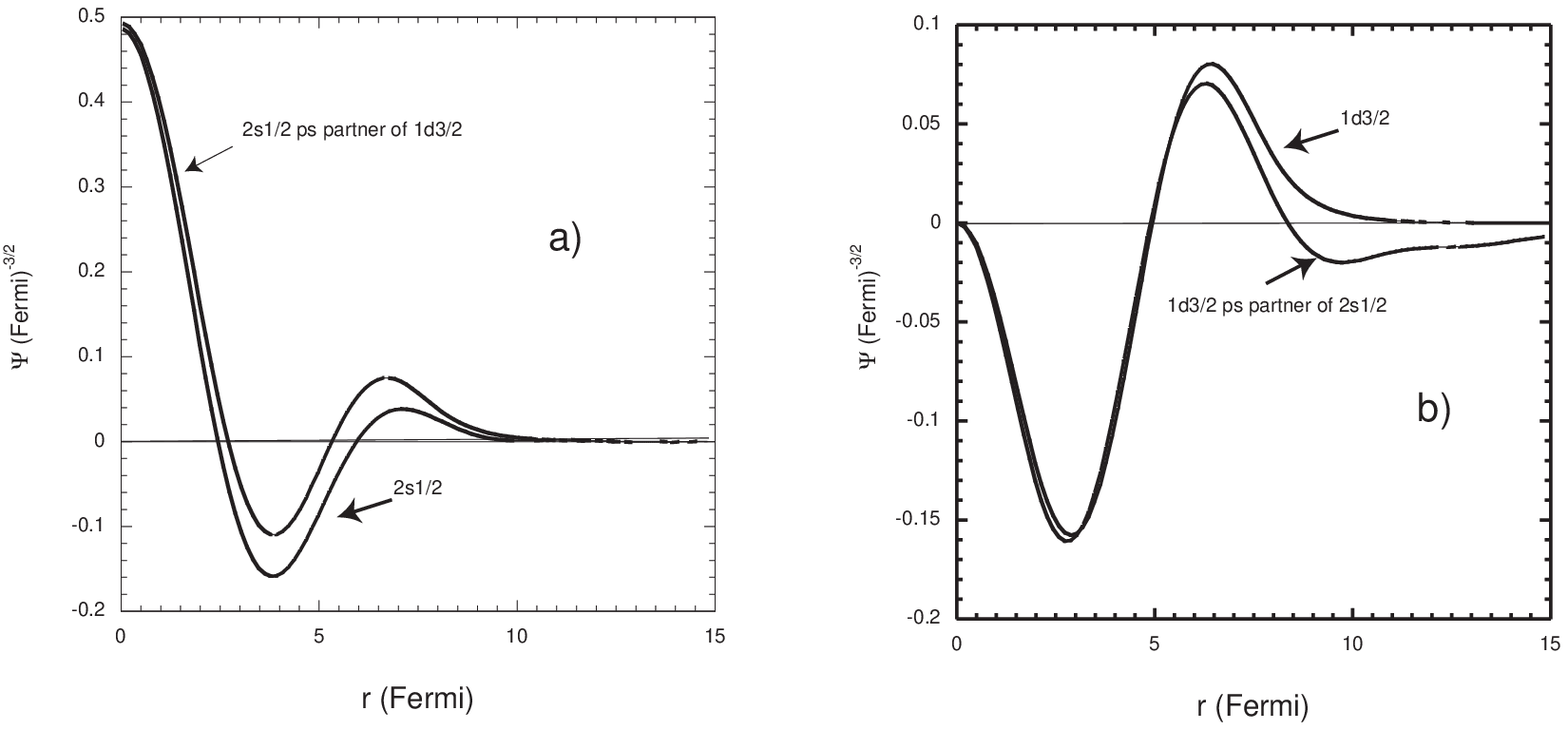 scaled 1000}
\end{center}
\noindent Figure 4. a) The $2s_{1/2}$ partner of $1d_{3/2}$ compared 
to the $2s_{1/2}$ eigenstate and b) the $1d_{3/2}$ partner of 
$2s_{1/2}$ compared to
the $1d_{3/2}$ eigenstate for relativistic mean field eigenfunctions 
for $^{208}Pb$ \cite{mad2}.\\
\end{figure}

\section{ QCD SUM RULES}

Applying QCD sum rules in nuclear matter \cite {furn}, the ratio of 
the scalar and vector self-energies were determined to be
${V_S \over V_V} \approx - {\sigma_N \over 8 m_q}$ where $\sigma_N $ 
is the sigma term which arises from the spontaneous breaking of
chiral symmetry \cite {cheng}.  For reasonable values of $\sigma_N $ 
and quark masses, this ratio is close to  -1. The implication of 
these results is
that chiral symmetry breaking is responsible for the scalar field 
being approximately equal in magnitude to the vector field, thereby 
producing
pseudospin symmetry.

\section{ NUCLEON - NUCLEUS SCATTERING}

The relativistic optical model scalar and vector potentials 
determined from nucleon-nucleus scattering are almost equal and 
opposite in sign
\cite{dave,bunny}. Since pseudosymmetry doesn't care if the 
potentials are complex, this symmetry may arise in nucleon-nucleus 
scattering
\cite{fred}. The pseudospin and spin symmetry breaking can be 
determined empirically if the polarization, $P$, and spin rotation 
function, $Q$, are both measured
as a function of the scattering angle \cite{gino4}. The scattering 
amplitude, $f$, for the elastic scattering
of a nucleon with momentum $k$ on a spin zero target in the spin 
representation is \cite{herman},
\begin{equation}
f = A(k,\theta ) +
B(k,\theta ) \mbox{\boldmath $\sigma $}{\cdot {\bf \hat n}}
\, .
\label{f}
\end{equation}
and in the pseudospin representation is
\begin{equation}
f = \tilde {A}(k,\theta ) +
\tilde {B}(k,\theta ) \mbox{\boldmath $\tilde {\sigma} $}{\cdot {\bf \hat n}}
\, .
\label{fps}
\end{equation}
The ratio, $\left |\frac{B}{A}\right |^2$, measures the amount of 
spin breaking and is given by  \cite{leeb},
\begin{equation}
\left |\frac{B}{A}\right |^2 =\frac{4}{2+2\sqrt{1-P^2-Q^2}-P^2-Q^2} 
\left[ \left( \frac{P}{2} \right)^2 +
\left( \frac{Q}{2} \right)^2 \right]
\label{Rs2}
\end{equation}
whereas $\left |\frac{\tilde B}{\tilde A}\right |^2$ measures the 
amount of pseudospin breaking,
\begin{equation}
\left |\frac{\tilde B}{\tilde A}\right |^2 = \frac{\tan^2(\theta ) - 
Q\tan (\theta )
+ \left |\frac{B}{A}\right |^2(1-Q\tan (\theta ))}
{1+Q\tan (\theta ) +\left |\frac{B}{A}\right |^2 (\tan^2 (\theta ) + 
Q\tan (\theta ))}
\, .
\label{Rps2}
\end{equation}
In Figure 5 the square of the ratio of the pseudospin dependent amplitude to
the pseudospin independent amplitude, $({\tilde B}/{\tilde A})^2$, 
and the square of the ratio of the spin dependent amplitude to the
spin independent amplitude, $ (B/ A)^2$, are plotted 
\cite{gino4,leeb} for 800 MeV proton scattering \cite {john} on 
$^{208}Pb$. The pseudospin spin breaking is
at most of the order of 10\%, a factor of three lower than the spin 
breaking. On the other hand low energy proton scattering indicates 
that pseudospin is badly broken \cite{fred}.
\begin{figure}[h]
\begin{center}
\HideDisplacementBoxes
\BoxedEPSF{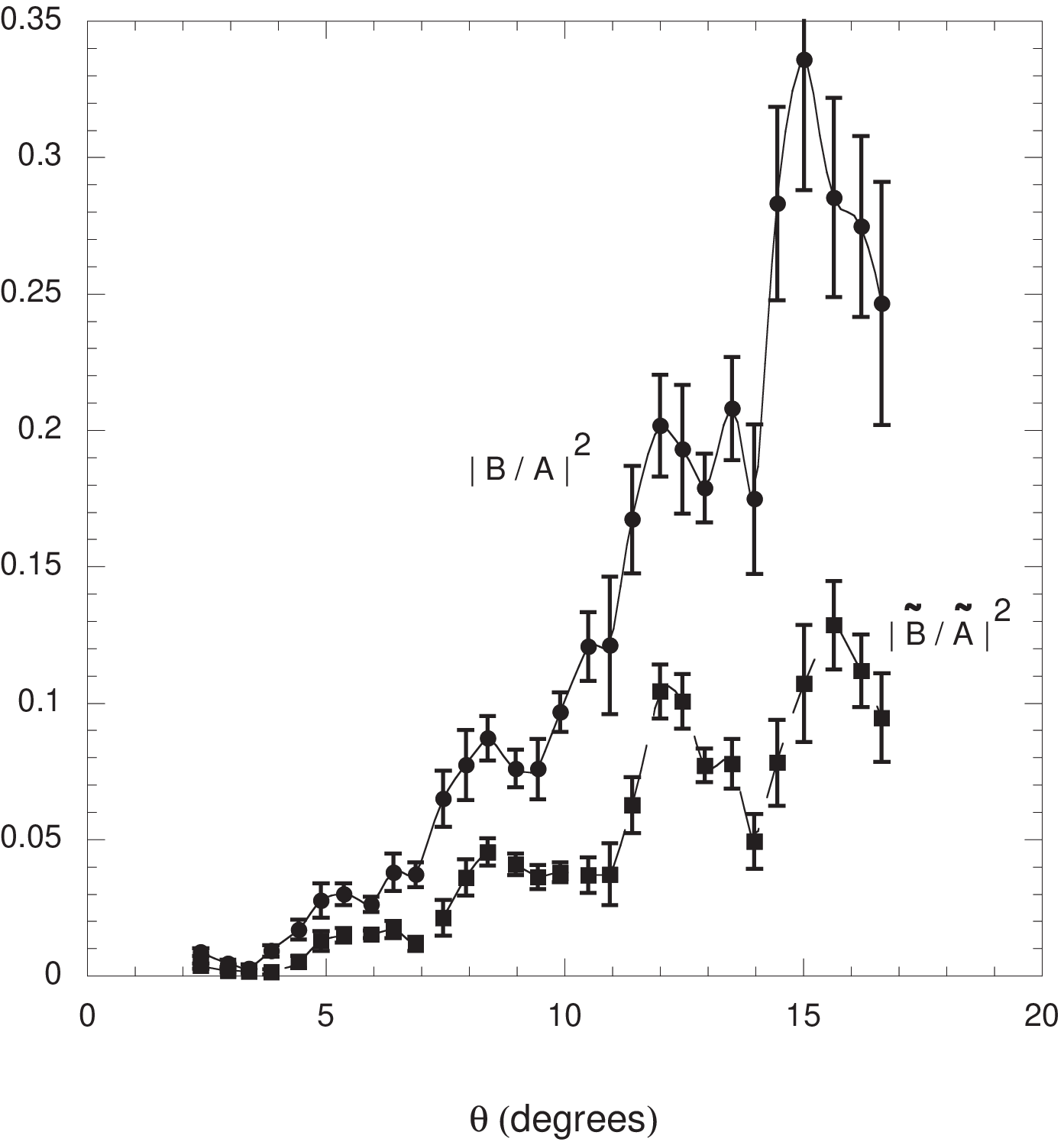 scaled 700}
\end{center}
\noindent Figure 5. The spin (dots), $\left |\frac{B}{A}\right |^2$, 
and pseudospin (squares), $\left |\frac{\tilde B}{\tilde A}\right 
|^2$, breaking probabilities are plotted
versus the scattering angle  \cite{gino4,leeb} for 800 MeV proton 
scattering \cite {john} on $^{208}Pb$.\\
\end{figure}

\section{SUMMARY}

We have shown that pseudospin symmetry is an SU(2) symmetry of the 
Dirac Hamiltonian with mean fields, $V_S = - V_V + 
constant$. This symmetry predicts that the eigenfunctions of the 
Dirac Hamiltonian will be pseudospin doublets
connected by the SU(2) generators. In nuclei, pseudospin symmetry 
must be broken since nuclei are not be bound in this limit. However
relativistic mean field approximations of 
relativistic nuclear field theories and relativistic nuclear 
Lagrangians with zero range interactions give 
scalar and vector mean field potentials in nuclei, $V_S 
\approx - V_V$.  We have shown that the pseudospin partners of the eigenstates of these
realistic relativistic Dirac Hamiltonians are themselves 
approximate eigenstates of the same Dirac Hamiltonian. Furthermore, since the 
realistic relativistic mean field calculations overestimate the energy splitting 
between doublets compared to their measured values, these same calculations may overestimate 
the difference 
between the experimental eigenfunctions. 

Another consequence of pseudospin symmetry is that, if the magnetic moments
of the states in the doublet are known, the ``$\ell$'' forbidden 
magnetic dipole transitions can be predicted. Similar 
relationships for Gamow-Teller transitions between states in the doublet 
hold as well.

Pseudospin symmetry has been shown to be
approximately conserved in medium energy nucleon-nucleus scattering from nuclei.

In the future these issues will be investigated in deformed nuclei as well.

Finally, pseudospin symmetry has been linked via QCD sum rules to chiral symmetry 
breaking in nuclei. This suggest a more fundamental significance which needs
to be explored. Recently much effort has been expended in deriving effective field 
theories based on QCD and chiral perturbation theory \cite {georgi}. However these effective field 
theories are generally non-relativistic field theories. Perhaps the lesson of pseudospin symmetry
is that relativistic effective
theories should be derived as well, at least as an intermediate step towards non-relativistic
field theories, in order to understand qualitative aspects about the physics of nuclei which may be missed
by non-relativistic theories. 

\section*{Acknowledgements}
This work was supported by the
United States Department of Energy under contract
W-7405-ENG-36. A. Leviatan, D. G. Madland, and P. von Neumann - Cosel 
collaborated on different aspects of the work reported in this 
survey. \\


\end{document}